\newcommand{\cm}{{~\rm cm}}
\newcommand{\s}{{~\rm s}}
\newcommand{\km}{{~\rm km}}

\newcommand{\K}{{~\rm K}}

\newcommand{\yr}{{~\rm yr}}

\newcommand{\AU}{{~\rm AU}}

\newcommand{\keV}{{~\rm keV}}
\newcommand{\MeV}{{~\rm MeV}}

\documentclass[12pt,preprint]{aastex}
\usepackage{color}      

\shortauthors{Sternberg \& Soker}

\begin{document}

\title{COMPARING SHOCKS IN PLANETARY NEBULAE WITH THE SOLAR WIND TERMINATION SHOCK}

\author{Noam Soker\altaffilmark{1}, Roi Rahin\altaffilmark{1}, Ehud Behar\altaffilmark{1},
and Joel H. Kastner\altaffilmark{2}}

\altaffiltext{1}{Department of Physics,
Technion$-$Israel Institute of Technology, Haifa 32000, Israel;
soker@physics.technion.ac.il}
\altaffiltext{1}{Center for Imaging Science, Rochester Institute of Technology,
Rochester, NY 14623-5604, USA}

\begin{abstract}
  We show that suprathermal particles, termed pick-up ions (PUIs),
  might reduce the postshock temperature of the fast wind and jets in
  some planetary nebulae (PNs) and in symbiotic systems.  The goal is
  to explain the finding that the temperature of the ``hot bubble'' formed
  by the post-shock gas in some PNs and symbiotic nebulae is lower,
  sometimes by more than an order of magnitude, than the value
  expected from simple hydrodynamical calculations.
  Although various explanations have been proposed, there is as
  yet no preferred solution for this ``low temperature
  problem.''  PUIs have been invoked to explain the low temperature
  behind the termination shock of the solar wind. While in the case of
  the solar wind the neutral atoms that turn into PUIs penetrate the
  pre-shock solar wind region from the interstellar medium (ISM), in
  PNs the PUI source is more likely slowly moving clumps embedded in the fast
  wind or jets.  These clumps are formed by instabilities or from
  backflowing cold gas.  Our estimates indicate that in young PNs
  these PUIs will thermalize before leaving the system.  Only in older
  PNs whose sizes exceed $\sim 5000 \AU$ and for which the fast-wind mass loss
  rate is $\dot M_w \la 10^{-7} M_\odot \yr^{-1}$ do we expect the
  PUIs to be an efficient carrier of energy out of the postshock
  region (the hot bubble).
\end{abstract}


\section{INTRODUCTION}
\label{sec:intro}

Central stars of young planetary nebulae (PNs) blow a fast wind that
collides with the dense PN shell and becomes a thermal X-ray source
(Volk \& Kwok 1985).  The postshock region of the fast wind is termed
the hot bubble.  About a third of PNs targeted by contemporary X-ray
satellite observatories (either {\it Chandra} or {\it XMM}) have been
shown to have extended X-ray emission (Kastner 2007). The other PNs
either posses a hot bubble but their emission is below detection
threshold, or the gas in the hot bubble escapes through a hole in the
dense shell that was punctured by a jet (or jets).  This last process
deserves further study.

The simplest and most straightforward estimate of the expected hot bubble temperature
is given by the Rankine-Hugoniot jump condition for the fast wind shock
\begin{equation}
T_{\rm fw} = \frac{3}{16} \frac {\mu m_H v_f^2}{k}
= 1.4 \times 10^7 \left( \frac{v_f}{1000 \km \s^{-1}} \right)^2 \K,
\label{eq:TH}
\end{equation}
where $v_f$ is the fast wind velocity, and the other symbols have
their usual meaning.  {{{  This expression is widely
      employed in describing astrophysical shocks in general and PN
      wind shocks in particular (e.g., Zhekov \& Perinotto 1996; Soker
      \& Kastner 2003, and references therein). }}} However, the
observed temperatures, $T_{\rm H}$, of the extended X-ray emission
sources in nearly all PNs (Kastner et al. 2000, 2001, 2003, 2008; Chu
et al. 2001; Guerrero et al. 2002, 2005; Sahai et al. 2003; Montez et
al. 2005; Gruendl et al. 2006) are lower than the simple estimates
obtained from Eq.~\ref{eq:TH}, with $T_{\rm H} \sim 1-3 \times 10^{6}
\K$ (Kastner 2007; Kastner et al. 2008).  The ratio of the temperature
given by equation (\ref{eq:TH}) to the observed temperature ($T_{\rm
  H} \sim 1-3 \times 10^{6} \K$) ranges from $T_{\rm fw}/T_{\rm H}
\sim 1.5$ (for one object, NGC~2392, adopting $v_f \simeq 400 \km
\s^{-1}$ [Tinkler \& Lamers 2002] and a revised estimate of $T_{\rm H}
= 1.5 \times 10^{6} \K$; Montez 2010, PhD thesis, RIT, in prep) to
$\sim200$ (NGC 7026, with $v_f \simeq 3500 \km \s^{-1}$ and $T_{\rm H}
= 1.1 \times 10^{6} \K$), with typical values of $T_{\rm fw}/T_{\rm H}
\sim10$.  For the one case in which the values of $T_{\rm fw}$ and
$T_{\rm H}$ might be consistent, NGC~2392, Tinkler \& Lamers (2002)
and Guerrero et al. (2010) report an anomalously slow wind speed of
only $v_f \simeq 400 \km \s^{-1}$. They further comment that the
central star (CSPN) of NGC 2392 is unusual among CSPN blowing winds.

We term the evidence that the X-ray emitting gas in PNs
possesses a temperature lower than that given by equation
(\ref{eq:TH}), $T_{\rm H} \la 0.1 T_{\rm fw}$ the {\it low temperature
  problem.}  Three different processes --- possibly acting in
combination --- were proposed to account for this low temperature
problem in the past:
\begin{enumerate}
\item  \emph{Heat conduction. }
The hot ($T> 10^7 \K$) post-shock gas is cooled via heat conduction to the cooler
nebular gas (Soker 1994; Zhekov \& Perinotto 1996; Steffen et al. 2008), or
via mixing with the cooler gas (Chu et al. 1997) enhanced by instabilities
(Stute \& Sahai 2006).
However, Yu et al. (2009) have shown that the composition of the X-ray-emitting
plasma in the hot bubble of BD +303639 is indistinguishable from that of the
CSPN fast wind.
There also exist preliminary indications that there is a sharp temperature jump
between the hot bubble and the cooler nebular gas (Nordon et al. 2009).
Hence it appears magnetic fields might inhibit heat conduction (Stute \& Sahai 2006).
\item \emph{Intermediate fast wind.}
In this solution to the low temperature problem the X-ray emitting gas comes mainly
from a slower moderate-velocity wind of $v_f \sim 500 \km \s^{-1}$ blown by the central
star during the post-AGB phase (Soker \& Kastner 2003; Akashi et al. 2006, 2007).
\item \emph{Hot lobes formed by jets.}  The X-ray emitting gas comes
  mainly from two opposite jets or collimated fast winds (CFW; Soker
  \& Kastner 2003; Akashi et al. 2008), expanding with velocities of
  $\sim 300-700 \km \s^{-1}$. The jets are blown by a companion
  accreting mass from the AGB or post-AGB star. In this scenario,
  which is most obviously applicable to objects such as NGC 7027 ---
  in which the X-ray morphology closely resembles that of
  high-velocity, collimated flows imaged in the infrared (Cox et al.\ 2002)--- the X-ray properties are tightly
  connected to the shaping mechanism of the nebula.
\end{enumerate}

As noted, some of these processes can coexist.  For example, both the
double opposite jets that have shaped the PN and the post-AGB
(intermediate) central star fast wind might occur one after the other
in a PN. Heat conduction and mixing between hot lobes and the nebular
gas might then occur, further lowering the X-ray temperature below
that expected given the present CSPN wind speed.

The low temperature problem is also evident in the case of some
symbiotic nebulae (some formed by symbiotic novae) that possess jets.
In the recurrent nova RS Ophiuchi the inferred jet velocity is $\sim
6,000 \km \s^{-1}$, for which equation (\ref{eq:TH}) gives a
temperature of $\sim 5 \times 10^8 \K$, while the inferred jet X-ray
temperature is only $\sim 10^7 \K$ (Luna et al. 2009).  The low
temperature problem also might exist in the case of the symbiotic nova
R Aquarii, where the X-ray temperature of $1.7\times 10^6 \K$ (Kellogg
et al. 2007) is much below the expected temperature of $8 \times 10^6
\K$.  If the underlying low temperature problem in these systems is
the same as in PNs, then the explanation cannot be an intermediate
fast wind, and we would need to appeal to one or both of the other
processes described above.

As there is no consensus yet on the mechanism that explains the low
temperature problem, in this paper we explore whether \emph{pick-up
  ions} (PUIs) can solve the low temperature problem in PNs.  In this
process, initially slowly moving ions embedded (but not moving with)
the fast wind are rapidly picked up by the fast wind and gain
super-thermal energy relative to the wind; hence the term PUIs.  In
the shock wave the PUIs gain much more energy than the thermal
particles and, hence, collectively act as a heat sink within the
post-shock gas. If the number of PUIs is high
enough, they can contain most of the post-shock energy and, in so
doing, substantially lower the post-shock temperature of the thermal
gas.

Measurements by the Voyager 2 spacecraft show {{{
      unexpected plasma properties.  For example, these measurements
      indicate that the flow is still supersonic with respect to the
      thermal ions downstream of the termination shock of the solar
      wind (Richardson et al. 2008; Decker et al. 2008).  Richardson
      et al. (2008) and Decker et al. (2008) conclude that most of the
      solar wind energy is transferred to PUIs or other
      energetic particles both upstream of and at the termination
      shock. }}} It is these unique, {\it in situ} measurements of a
wind shock that serve as the primary motivation for the work presented
here. However, we note that PUIs also may represent a significant
cooling mechanism in supernova remnants (SNR; e.g., Ohira \& Takahara
2010 for a recent paper and more references).  In SNRs the PUIs play a
role mainly in the forward shock that runs ahead of the interaction
region (Ferrand et al. 2010), while for the solar wind --- and, as we
demonstrate below, in PNs --- the PUIs play a role in the reverse
shock that runs into the stellar wind.  In these latter systems the
forward shock is a weak shock (i.e., has a low Mach number).  While
the physics of the shocks in SNRs and in PNs is the same, we argue
below that the way the PUIs enter the pre-shock wind region is
different.

Our study of the formation and behavior of possible PUIs in PNs is
organized as follows.  In section 2 we summarize the main relevant
results of the PUIs in the solar wind.  In section 3 we compare the
properties of the solar termination shock to those of the fast wind in
PNs, and in section 4 we study the constraints on the flow properties
for PUIs to play a role in PNs.  In section 5 we summarize our main
results.

\section{THE ROLE OF PICKUP IONS IN THE SOLAR WIND TERMINATION SHOCK}
\label{sec:solar}

{{{  Many different ingredients and processes that occur
      in the solar wind and its termination shock are not relevant to
      PNs.  As we will see later, such differences are due to the fact
      that the mean free path of neutrals and PUIs in PNs is much
      shorter than in the solar wind, and the source of PUIs in PNs is
      different than that in the solar wind-ISM interaction.
Hence, in
      this section we describe the properties of solar wind and
      termination shock PUIs that are appropriate to consider in the
      context of PN wind shocks. For a complete description of solar
      wind and termination shock processes --- including references to
      additional papers concerning the solar wind, its termination
      shock, and its interaction with the ISM --- the reader is
      directed to Zank et al. (2010). }}}

\begin{figure}
\includegraphics[scale=0.6]{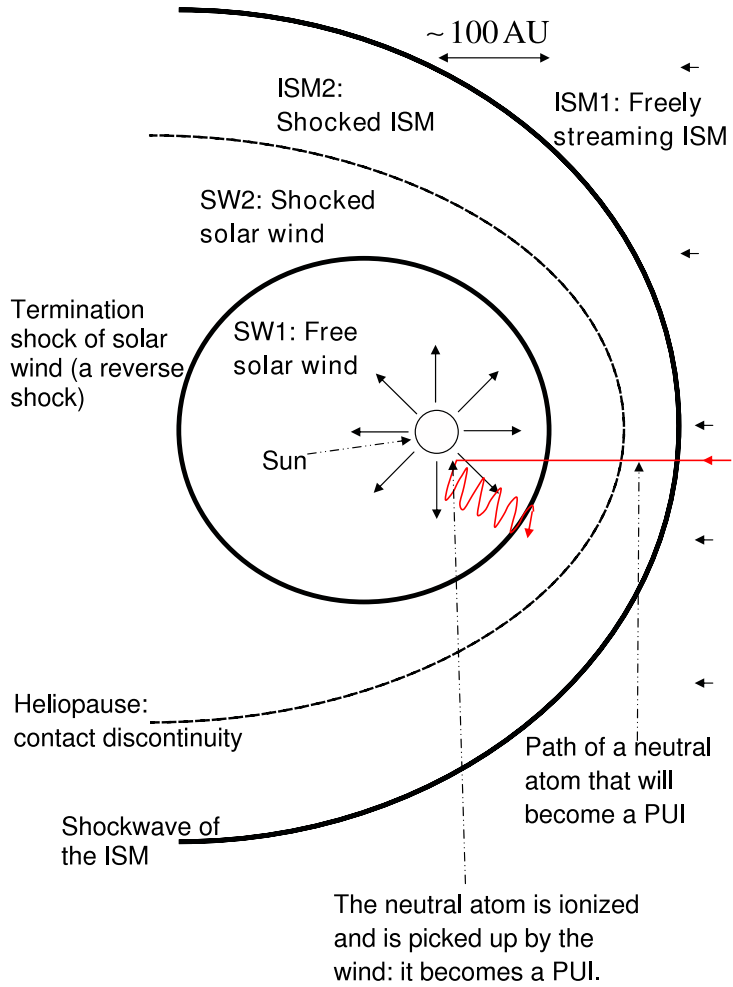}
\caption{Schematic (not to scale) drawing of the interaction of the
  solar wind with the ISM.  The ISM flows from right to left.  The
  trajectory of the neutral ISM atom that turns into a PUI (see text)
  is schematically depicted by a line (in red).}
\label{fig:solar}
\end{figure}

The schematic solar wind flow structure is drawn in
Fig. \ref{fig:solar}.  As shown schematically in the diagram (and will
be quantitatively derived later in the paper), neutral atoms in the
ISM have a very long mean free path to collision, and can penetrate to
several AUs from the sun {{{  (see review in Zank
      1999). }}} There they are ionized and instantaneously picked-up
by the magnetic field in the solar wind; these particles therefore
become PUIs.  {{{  After being picked up, the PUIs
      experience scattering and isotropization by either ambient or
      self-generated low-frequency electromagnetic fluctuations in the
      solar wind plasma (Zank 1999). }}} {{{  As they are
      now isotropized, the bulk velocity of the PUIs is that of the
      solar wind.  Namely, the average location of the PUIs, as they
      gyrate about the interplanetary magnetic field, comoves with the
      solar wind; but their temporary speed at each moment relative to
      that location }}} is about equal to the solar wind speed
{{{  (e.g., Figs. 3.8 and 3.13. in Zank 1999), as
      they have high kinetic energy ($\sim 1 \keV$).  In essence, the
      solar wind protons form a relatively cold ``core'' about
      which is superimposed a dilute halo of energetic PUIs.
      The PUIs suffer adiabatic cooling and energy diffusion, and
      their distribution with energy changes with distance from the
      Sun (e.g., Fig.\ 3.9 in Zank 1999). They also can heat the
      cold core proton distribution of the solar wind (Zank 1999).  As
      we will see later, in PNs we require the PUIs to form close to
      the termination shock, and these processes are not
      significant. }}} When they cross the reverse shock they gain
relatively more energy than the thermal particles.

The existence of PUIs in the solar wind was suggested a long time ago
to explain the overabundance of O and N in cosmic rays between
5-30~MeV (Fisk et al. 1974; Hovestadt et al. 1973; {{{
      also see review in Zank 1999). }}} This idea {{{
      (Zank et al. 1996) }}} was recently revisited {{{
      (Richardson 2008)}}} in the context of Voyager 2's measurement
for the temperature of the solar wind termination shock, which was
much lower than the hydrodynamic value expected from
equation~\ref{eq:TH}(Richardson et a. 2008).  Specifically, at the
time when Voyager 2 crossed the termination shock, August 30 to
September 1 2007, the post-shock temperature was $\sim 10^5 \K$ ---
not $\sim 10^6 \K$, as predicted by equation~\ref{eq:TH} --- and the
post-shock flow remained supersonic (Richardson et a. 2008).  The
explanation for this deviation from the simple hydrodynamic post-shock
conditions is that most of the energy in the post shock region (region
SW2 in Fig. \ref{fig:solar}) is carried out by suprathermal particles,
i.e., PUIs .  {{{  The pickup ions make up $\sim 20 \%$
      of the sheath plasma and have energies of $\sim 6 \keV$ in the
      heliosheath region SW2 (Richardson 2008), as compared with $\sim
      1 \keV$ per nucleon in the pre-shock region (SW1 in
      Fig. \ref{fig:solar}). }}} {{{  Indeed, the
      presence of these energetic particles was deduced from}}}
{{{  low-energy ions measured by }}} Voyager 2 (Decker et
al. 2008), and have been theoretically shown to explain the measured
post-shock solar wind temperature (e.g., Fahr \& Chalov 2008; Wu et
al. 2009; {{{  Zank wt al. 2010). }}}

The chain of the physical processes in the solar wind and its
termination shock is as follows (Fisk et al. 1974; {{{
      Zank 1999; Zank et al. 1996, 2010). }}} Because of their long
mean free path, neutral atoms in the interstellar medium (ISM; region
ISM1 in Fig.~\ref{fig:solar}) are not influenced by the collisionless
forward shock that is formed as a result of the relative motion of the
sun and the ISM.  Moreover, {{{  many of the }}} neutral
atoms are neither influenced by the shocked solar wind (region SW2)
nor by the outer region of the pre-shock solar wind (SW1;
{{{  for a detailed description see Zank 1999 and
      Zank et al. 2010). }}} Only when they reach the inner regions
close to the sun, $\sim 1-10 \AU$, are they {{{
      efficiently }}} ionized by either charge exchange with the solar
wind ions or by the solar UV flux (e.g., Vasyliunas \& Siscoe 1976).
Subsequently, the now-charged particles are picked up by the
magnetized solar wind.  Their kinetic energy per unit mass relative to
the solar wind is given by
\begin{equation}
e_{\rm PUI1} \sim (1/2) v_{\rm sw}^2 \simeq 1 \keV ~{\rm nuc}^{-1},
\label{eq:epui1}
\end{equation}
where $v_{\rm sw} \sim 500 \km \s^{-1}$ is the solar wind velocity.
{{{  With this high average particle energy the PUI
      pressure dominates in the outer heliosphere (region SW1) (e.g.,
      Zank et al. 2010 and references therein). }}}
As they cross the termination shock from SW! to SW2, the PUIs can acquire energies of up to
$\sim 100 \MeV {\rm nuc}^{-1}$ (Ellison et al. 1999); {{{  but only a very
      small number of PUIs reach such high energies. }}} The average
gain of energy by the PUIs is (Fahr \& Chalov 2008; {{{
      eq. 8 in Zank et al. 2010) }}}
\begin{equation}
\Delta E_{\rm PUI} \simeq E_{\rm PUI1} \left( s^2 -1 \right) ,
\label{eq:dpui}
\end{equation}
where $E_{\rm PUI1}$ is the pre-shock energy of the PUIs, while $s=\rho_2/\rho_1$
is the compression ratio, and throughout this paper it is assumed that the
adiabatic index is $\gamma=5/3$.
Because of the PUIs, the compression factor of even a strong shock is
$<4$ {{{  (Zank 1999).  The process of formation of PUIs
      results in a reduction in the bulk kinetic energy of the wind
      and the heating of the gas, both of which reduce the upstream
      Mach number (Zank 1999). }}} The shock compression ratio found
by Voyager 2 is $\sim 2.4$ (Richardson et al. 2008).
{{{  We note that Stone et al. (1996) deduced a
      compression ratio of $2.63 \pm 0.14$ in the solar termination
      shock in 1994. They based their estimate on the energy spectra
      of anomalous cosmic rays measured by the Voyager and Pioneer
      spacecraft during 1992-1994. }}}

Overall, for the PUIs to carry
$90 \%$ of the postshock energy, the energy of the PUIs before the
shock should be $E_{\rm PUI1} \simeq 0.2 E_{t1}$, where $E_{t1}$ is
the total energy of the preshock wind, which is practically the
kinetic energy of the pre-shock wind.  The energy of the pre-shock
PUIs per unit mass is given by equation (\ref{eq:epui1}); that of the
wind's kinetic energy is about the same. Therefore, by equation
(\ref{eq:dpui}), the number of the PUIs should be $\sim 0.2$ times
that of the wind particles.  Indeed, using the PUIs flux at the nose
of the heliosphere calculated by Cummings \& Stone (1996), Ellison et
al. (1999) find the density of the PUIs at the termination shock to be
$\sim 0.2$ times that of the thermal solar wind.  {{{
      Similar values were found by Richardson (2008) }}} and Wu et
al. 2009.

\section{THE SOLAR WIND SHOCK VS.\ THE FAST WIND SHOCK IN PLANETARY NEBULAE}
\label{sec:fast}

\begin{figure}
\includegraphics[scale=0.6]{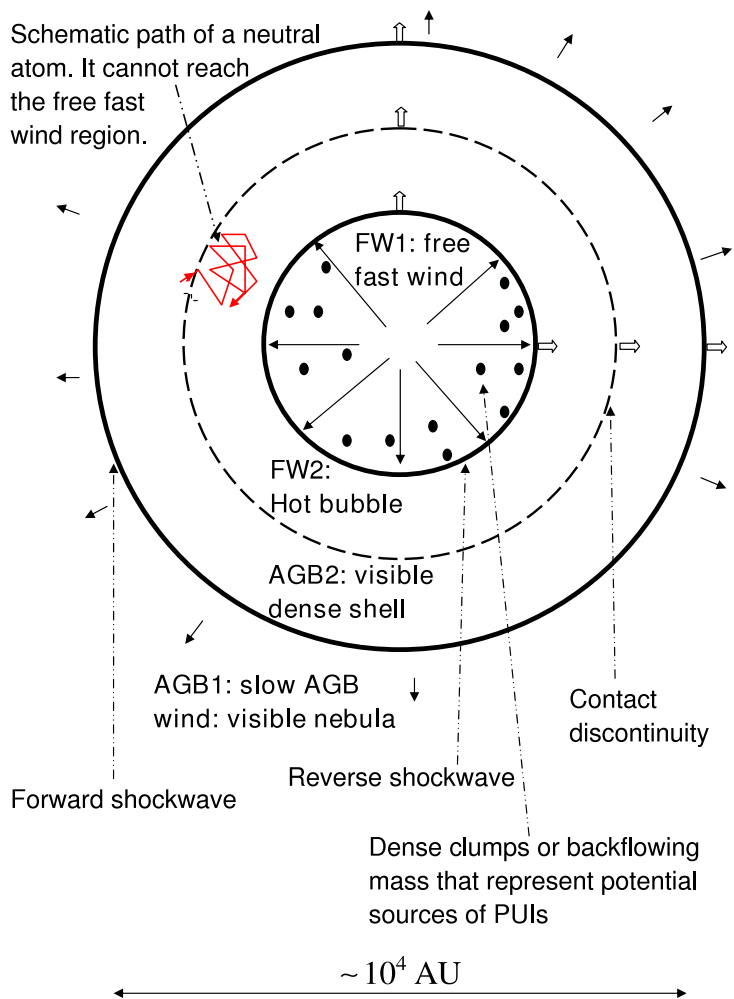}
\caption{Schematic (not to scale) drawing of the interaction of the wind blown by the
central star of the planetary nebula (CSPN) with the slow wind blown by the progenitor AGB star.
Arrows show the flow, while double arrow indicate that the entire structure expands in the radial direction.
Because of the much higher density than in the case of the solar wind, neutral atoms cannot
penetrate the shocked fast wind region (FW2) and reach the pre-shock fast wind region; this
is depicted by the zig-zag (red) line.
PUIs that are formed close to the CSPN will lose their energy before reaching the reverse shock.
Therefore, any significant source of PUIs must originate in slowly moving clumps inner to,
and close to, the reverse shock.}
\label{fig:pn}
\end{figure}

With the physics of the solar wind termination shock in mind, we turn to
examine the termination shock of the fast wind blown by the central star
of PNs (CSPN).
In Figure \ref{fig:pn} we present the schematic flow structure, while in Table 1 we
compare several properties of PNs with those of the solar wind interaction with the ISM.
We expect that more can be learned about PNs by comparison to the solar wind interaction
with the ISM, but in this work, we limit ourselves to the role of PUIs.
\begin{table}
\caption{Comparison of the Solar Wind and PN Fast Winds}
\begin{tabular}{|l|c|c|}
\hline
Property              &         Solar wind        &           Winds in PNs     \\
\hline
Wind velocity (${\rm km} \s^{-1}$)        & $v_{\rm sw} \simeq 500$        & $v_{f} \sim 500-1500$ \\
Mass loss rate ($M_\odot \yr^{-1})$       & $\dot M_{\rm sw}\sim 3 \times 10^{-14}$       & $\dot M_{f} \sim 10^{-6}-10^{-8}$ \\
Shock position $r_s$ (AU)                 & $r_{\rm sw} \sim 100$          & $r_{\rm sf}\sim 1000 - 10^4$ \\
Wind density at $r_s$ $({\rm g} \cm^{-3})$& $\rho_{\rm sw} \sim 10^{-27}$  & $\rho_{\rm fw} \sim 10^{-26} - 10^{-22} $ \\
External region  $({\rm g} \cm^{-3})$     & $\rho_{\rm ISM} \sim 10^{-24}$ & $\rho_{\rm AGB} \sim 10^{-20}- 10^{-17}$  \\
Mean free path$^{(1)}$ $\lambda_{\rm N}$~(AU)& $\lambda_{\rm SW2} \sim 10^5$ &  $\lambda_{\rm FW2} \sim 1-10^4 $  \\
$\lambda/r_s$                         &    $\sim 10^3$                 &  $10^{-3}-1$   \\
PUI stopping time$^{(2)}$ (years) $t_s$         &  $\sim 10-100$                           &  $\sim 1-10$   \\
Hot region age$^{(3)}$ (years)   $t_{\rm age}$ &  $\sim 1$                              &  $\sim 10^3$  \\
Preshock flow time (years)       $t_f$         &  $\sim 1$                              &  $\sim 10-100$  \\
$t_s/t_f$                                      &  $\sim 10-100$                         &  $\sim 0.01-1$  \\
\hline
Ionization structure                  & The ISM is mostly     & During the PN phase   \\
of the outer region                   & neutral.              & the entire outer  \\
                                      &                       & region is ionized.    \\
\hline
Source of PUIs                        & ISM outside the     & Backflowing mass or  \\
                                      & shocked regions.    & slow clumps embedded  \\
                                      &                     & in the fast wind.  \\
\hline
\end{tabular}

{NOTES: (1) The mean free path is for neutral particles moving through the shocked region
of the solar wind and fast wind, respectively. (2) The stopping time of the PUIs in the pre-shock
(free wind) region. In the hot bubble the stopping time is longer.
(3) In the solar wind the temperature is measured near the termination shock,
and the age of the shocked wind is very short. In PNs the hot shocked gas remains in the hot bubble
for $>1000 \yr$.  }
\end{table}

As we show here and in Sec.~4, if PUIs exist and represent an
important temperature-regulating process in PNs, then perhaps the most
significant difference between the solar wind and PNs is the source of
the PUIs.  The source of the PUIs in the solar wind are neutral atoms
that penetrate from the ISM.  The mean free path for the neutral atom
to collide with the shocked solar wind is large, and they can easily
reach the pre-shock solar wind. The situation in the case of PNs is
the opposite.  We take a compression ratio of $2.4$ as found in the
solar termination wind (Richardson et al. 2008), and find the
postshock ion density in a spherical wind to be
\begin{equation}
n_{2i}({\rm post}) = 1  
\left( \frac{\dot M_w}{10^{-7} M_\odot \yr^{-1}} \right)
\left( \frac{r_s}{5000 \AU} \right)^{-2}
\left( \frac{v_w}{1000 \km\s^{-1}} \right)^{-1} \cm^{-3},
\label{eq:n2i}
\end{equation}
where $r_s$ is the shock radial distance, $v_w$ is the velocity of the central wind
($v_{\rm sw}$ for the sun and $v_{f}$ for PNs),  and $\dot M_w$ is the mass loss rate of the wind.

In Table 1 we compare the wind parameters of the solar wind with those
of PNs. It is apparent that the largest difference between the two
lies in the fast wind mass loss rate. While the shock in PNs can be at
$\sim 1000-10^4 \AU$,which is only 10-100 times the shock radius of
the solar wind, the mass loss rate is $\sim 10^{-8}- 10^{-6} M_\odot
\yr^{-1}$, which is more than five orders of magnitude larger than
that of the solar wind. The wind velocities are similar. For a
collision cross section of $\sigma \simeq 3 \times 10^{-16} \cm^{-2}$
(Heng \& Sunyaev 2008 and references therein) the mean free path of
neutral atoms in the post-shock solar wind region is $\lambda_{\rm N}
\simeq 10^5 \AU \gg r_s$. For an ``optimistic'' case in which these
neutral atoms might penetrate deeply into the PN
$r_s \simeq 2 \times 10^4 \AU$, $\dot M_w = 4 \times 10^{-8} M_\odot \yr$
and ${v_w} \simeq 2000 \km\s^{-1}$
(values similar to those determined for NGC 6543; Kastner et al.
2008), we find $\lambda_{\rm N} \simeq 2 \times10^4 \AU \simeq r_s$. Even in this case, the mean
free path is not much larger than the shock radius. However, such a
large bubble is formed at a late stage, when the central star has
already ionized the entire PN dense shell (regions AGB1 and AGB2 in
Fig. \ref{fig:pn}), and there is no source of a large number of
neutral atoms that can penetrate through the shocked fast wind without
becoming ionized. Hence, for many (if not most) PNs the shock radius
is smaller and the mass loss rate is much higher, and therefore the
mean free path is much smaller than the shock radius. We conclude
that, unlike in the case of the solar wind, the PUIs are unlikely to
originate in the regions outside the shocked fast wind.

\section{PICK UP IONS (PUIs) IN PLANETARY NEBULAE}
\label{sec:PUI}

\subsection{The Source of the Pick-Up Ions (PUIs)}
\label{subsec:source}
By analogy with the solar wind, the fraction of PUIs in the fast wind
of PNs should be $0.05-0.3$ if PUIs represent an important cooling
mechanism for the post-shock gas.  Here, we consider in detail two
potential sources for the PUIs.  We first note that the ablation rate
of ions from Uranus-like planets that might orbited the progenitors of
some PNs is of the order of $\sim 10^{-14} M_\odot \yr^{-1}$ (Soker
1999). The rate will be lower for minor planets. Even if we consider
millions of minor planets associated with a PN progenitor star, this
rate is too low to to provide the PUI fraction necessary to explain
the low temperature problem.  In addition, these objects would reside
relatively close to the center of the PN, such that the PUIs will
reach (almost) equilibrium with the thermal gas before they reach the
shock wave, as we show in section \ref{subsec:path} below. As we see
below, clumps (knots) in PNs are much larger than planets, and might
contain enough mass; they are also more widely distributed within the
PN, and their ablation (evaporation) rate is much higher.
{{{  The formation of PUIs from clumps within PNs
      resembles in some aspects the formation of PUIs from comets (and
      even planets) in the solar system (regarding the potential
      cometary origin of solar wind PUIs see, e.g., Bzowski \&
      Kr{\'o}likowska 2005). }}}

Formation of neutral atoms in the hot bubble (post-shock gas) also
likely does not represent a potential significant source of PUIs. The
neutral atoms are not influenced by the magnetic field, and their mean
free path is $\lambda_N \sim 200 (n_e/1 \cm^{-3}) \AU$. This is the
region from where neutral atoms formed by recombination in the hot
bubble might reenter the preshock fast wind (FW1) region. However, in
a typical (evolved) PN, there are not enough of these neutral atoms.
The recombination time at the relevant temperature of $\sim 10^6 \K$ is
$\sim 5 \times 10^6 (n_e/1 \cm^{-3})^{-1} \yr$, which is too long to
supply the required neutral atom fraction of at least 0.05 of the total fast
wind particles.
If we consider the photoionization during the PN phase or the
expectation that only a fraction of the recombined atoms will diffuse
toward the upstream direction, then it is much less likely that
recombining neutrals can be an important source of PUIs in PNs.  At
very early PN stages the shock radius $r_s$ is very small and the
recombination time of the postshock fast wind is short, and there is
no significant ionizing radiation yet.  However, if recombination
proceeds fast enough to supply the neutral atoms, then the postshock
gas cools rapidly via this same recombination process, such that
there would be no low temperature problem in very young PNs.

\subsubsection{Slowly moving clumps.}
We discuss here a case where the source of the PUIs in PNs is slowly
moving clumps embedded in the fast wind. Whether such clumps start
neutral or ionized, once ions (mainly protons) are released by the
clumps, they would be picked up by the wind. Because of the large
velocity difference between the slowly moving clumps and the fast
wind, these ions behave like the PUIs in the solar wind. It is trivial
to estimate the total mass that would need to be evaporated from the
clumps to contribute sufficient PUIs to cool the postshock region: the
bulk of the mass in the fast wind is lost over the first $\sim 1000$
years, with a mass loss rate of $\sim 10^{-8} -10^{-6} M_\odot
\yr^{-1}$ (Kastner et al. 2008) and, hence, for a PUI fraction of 0.1,
the evaporated mass from the slowly moving clumps should be $\sim
10^{-6} - 10^{-4} M_\odot$.  For $>1000$ clumps, the mass in each
clump could be very small, $\la 10^{-7} M_\odot$. However, the clumps
would need to be distributed in the inner volume of the PN.

{{{  The best studied case of slowly moving ($\sim10$ km
      s$^{-1}$) clumps are the cometary globules (or knots) in the
      Helix nebula (NGC 7293). Although the origin of these clumps (or
      knots) remains uncertain, their presence is well established
      (O'Dell et al. 2007 and references therein).}}} The
characteristic mass of individual knots is estimated to be $\sim 1-5
\times 10^{-5} M_\odot$ (O'Dell \& Burkert 1997; Meaburn et al. 1992;
Huggins et al. 2002; Meixner et al. 2005). Their total number in the
inner region, {{{  which is engulfed in fast wind gas,}}}
is $\sim 10^4$ (O'Dell \& Handron 1996; Hora et al. 2006).  The total
mass in the knots of the Helix nebula is therefore about three orders
of magnitude above the required mass. The knot evaporation time scale
during the early PN phase is $\sim 10^5 \yr$ (Matsuura et
al. 2009). Therefore, the evaporation rate of slowly moving gas in the
Helix is $\sim 10^{-6} M_\odot \yr^{-1}$, about an order of magnitude
higher than required. For smaller clumps the evaporation rate would be
higher, and a mass of only $\sim 10^{-4} M_\odot$ might be sufficient
to reduce the post-shock temperature of the fast wind.

Although there are other PNs with cometary globules similar to those
observed in the Helix nebula {{{  (see, e.g., images of
      the Ring Nebula at the Hubble Heritage
      website\footnote{http://heritage.stsci.edu/1999/01/index.html}
      and of the narrow-waist bipolar PN NGC 6302 in Szyszka et
      al. 2009),}}} we caution that the Helix is not an ideal case to
consider for purposes of evaluating whether slow-moving clumps could
serve as a source of PUIs.  The Helix has no detectable diffuse (hot
bubble) X-ray emission (Guerrero et al. 2001) and, even at early PN
stages, its (apparent nearly pole-on bipolar) nebular geometry may not
have been conducive to the presence of such emission (Kastner et al.\
2008). It therefore remains to be determined whether the presence of
Helix-like clumps, either neutral (as in the Helix) or ionized, are a
common feature of PNs.

\subsubsection{Back-flowing material. }
Backflowing material might supply plenty of PUIs, in addition to the
slowly moving clumps.  Accretion of backflowing material during the
post-AGB phase was considered in several papers (Mathis \& Lamers
1992; Bujarrabal et al. 1998; Zijlstra et al. 2001; Soker 2001) to
explain other properties of PNs such as longer post-AGB evolution
time.  {{{  This type of accretion is driven by the
      gravity of the central star that acts efficiently on dense
      clumps (Soker 2001). The dense clumps are formed during the AGB
      phase, probably requiring the presence of a binary companion
      (Soker 2010).  Post AGB accretion rates of $\ga 10^{-7} M_\odot
      \yr^{-1}$ were considered by Soker (2001); this theoretical
      study}}} showed that conditions might exist for a total accreted
mass of $\sim 0.1-0.001 M_\odot$.  {{{  Before the PN
      phase the clump is neutral.  After the beginning of the PN
      phase, which is the phase relevant to us, the outer regions of
      the clumps are ionized and evaporated. These ions could be
      picked up by the fast wind. If a fraction of $\sim 10^{-3} -
      0.1$ of the accreted mass }}} were picked up by the fast wind,
this would be sufficient to substantially reduce the post-shock
temperature.

Backflow might persist to the PN phase (Frankowski \& Soker
2009). {{{  The backflow in this case is driven by the
      rapid increase in the pressure of the dense, slowly expanding
      main shell of the PN. The rapid increase in pressure occurs when
      the central star starts to ionize the nebula and the nebular
      temperature increases from $\la 10^3 \K$ to $10^4 \K$ in a short
      time.  A rapid increase in the pressure of the shell may also
      result in instabilities that form ionized clumps (Frankowski \&
      Soker 2009). In the PN phase the backflow rate is only $\sim
      10^{-9} - 10^{-6} M_\odot \yr^{-1}$ over thousands of years, and
      the total backflowing mass is $\sim 10^{-5} - 10^{-3} M_\odot$.
      This is sufficient for the PUI process to be important, }}}
because over thousands of years the mass loss rate of the fast wind
substantially declines, and so does the required PUI mass.

The backflowing gas fulfills the condition that the source of the PUIs
be near the reverse shock, and just interior to it, such that the PUIs
have no time to reach equilibrium with the thermal pre-shock fast wind
(see section \ref{subsec:path}).  According to Frankowski \& Soker
(2009) the backflowing gas is well protected as it falls within the
shocked fast wind zone. However, as clumps of gas cross the shock
front (as they fall toward the center), they would become subject to
the ram pressure of the fast wind. There they would be quickly
decelerated, and could be destroyed, so as to provide a source of PUIs.

\subsection{The stopping time of the PUIs}
\label{subsec:path}
Even if the PUIs originate in the inner region, there is an additional requirement for
PUIs to play a role in the low temperature problem.
The PUIs that are formed in the pre-shock fast wind in PNs should not reach thermal equilibrium with
the gas, neither before nor after passing through the shock.
More quantitatively, the stopping time $t_s$ of the PUIs needs to be longer than
the relevant flow time.
In the postshock region (the hot bubble) the relevant time is the PN age, $t_{\rm age} \sim 10^3 \yr$.
For PUIs in the pre-shock wind the relevant flow time is the flow time from their origin
to the shock.
If they are forming a distance $\Delta r$ inward to the shock, this time is
\begin{equation}
t_f \simeq 5           
\left( \frac{\Delta r_s}{0.2 r_s} \right)
\left( \frac{r}{5000 \AU} \right)
\left( \frac{v_w}{1000 \km \s^{-1}} \right)^{-1} \yr.
\label{eq:tf}
\end{equation}
This requires the PUIs to originate far from the central star, where the density drops ($n \propto r^{-2}$)
and in low mass loss rate PNs (as $n \propto \dot M$; see eq.~\ref{eq:n2i});
density and mass loss rate thus play a role in determining the stopping time (see below).

We can use Spitzer (1956) to estimate the stopping time (and hence the mean free path $\lambda _\mathrm{mfp}$)
of a high-speed PUI originating in the fast wind before it is stopped inside the shocked hot
bubble ($T \approx 10^6$ K, $n_e \approx 1$~cm$^{-3}$) and in the more extended cold nebula
($T \approx 10^4$~K, $n_e \approx 10^4 $~cm$^{-3}$).
In the pre-shock wind we take the wind temperature to be $\sim 10^4 \K$, where the ionizing radiation of
the CSPN will balance adiabatic and radiative loses.
Using Eq. (5-28) of Spitzer (1956), we write the typical time it would take a PUI to be stopped by
dynamical friction in the pre-shock wind
\begin{eqnarray}
t_s = \frac{kTM_+V_+}{4\pi e^4n Z^2 Z_+^2\ln \Lambda G(V_+/v_\mathrm{th})}
\qquad \qquad \qquad \qquad
\nonumber
\\
\simeq 1.8
\left( \frac{T}{10^4 \K} \right)
\left( \frac{M_+}{m_p} \right)
\left( \frac{V_+}{1000 \km \s^{-1}} \right)
\left( \frac{n}{1 \cm^{-3}} \right)^{-1}
\left( \frac{{\rm ln} \Lambda}{30} \right)^{-1}
\left( \frac{G}{0.2} \right)^{-1}
Z^{-2}Z^{-2}_+ \yr,
\label{eq:ts}
\end{eqnarray}
where $M_+$, $Z_+$, and $V_+$ are the PUI mass, charge, and velocity, respectively;
$T$, $n$, $Z$, and $v_\mathrm{th}$ are the temperature, density, charge, and thermal
velocity of the ambient gas.
{{{  Note that $V_+$, the thermal velocity of the PUIs relative to the bulk of the wind, is equal to the bulk velocity of the wind that picks up the PUIs. }}}
The Coulomb logarithm $\ln \Lambda \simeq 20 - 30$, while $G$ is a function
of the velocity ratio and is tabulated by Spitzer (1956).
The second line of Eq.~(\ref{eq:ts}) incorporates values appropriate for the fast PN wind.

In the pre-shock region the electron velocity is similar to the PUI speed relative to the wind,
and the electrons stop the PUIs, while in the hot bubble the PUIs are slowed down most efficiently by
ions.
The reason is as follows.
When the PUI velocity much exceeds the thermal velocities, $t_s \propto T^{3/2}$ and is
independent of $V_+$.
Indeed, the function $G$ attains its maximum value and, hence,
$t_S$ is a minimum, for $V_+ = \sqrt{2kT/m} $.
For a $10^3 \km \s^{-1}$ PUI this is approximately the case for the thermal ions
in the hot bubble and for the thermal electrons in the cold nebula.
Using equation (\ref{eq:n2i}) to find the density in the solar wind, and then plugging this value into equation
(\ref{eq:ts}), we find that the PUIs in the solar wind lose their energy
on time scales much longer than the flow time scale (see Table 1).

Comparing equations (\ref{eq:ts}) and (\ref{eq:tf}), with the aid of
equation (\ref{eq:n2i}), we find that for the PUIs not to lose their
suprathermal status before the shock, the mass loss rate of the fast
wind is required to be $\dot M _w \la 10^{-7} M_\odot \yr^{-1}$ and
the shock to be at $r_s \ga 5000 \AU$. Both of these conditions apply to
relatively evolved PNs, and hence are not a strong constraint, because the low
temperature problem arises for PNs for which the fast wind speed is $v \ga
10^3 \km \s^{-1}$.
Another constraint is that the PUIs source
be close to the shock  (\S 4.1.2), both for the flow time $t_f$ to be short, and
for the densities at the sites of PUI origin not to be too high.

In the postshock region the temperature is {{{  typically
      observed to be }}} $\sim 3 \times 10^6 \K$ {{{
      (Kastner 2007; Kastner et al. 2008), }}} and the PUIs have been
accelerated to $v \ga 2000 \km \s^{-1}$.  {{{  This
      velocity comes from the energy gain in the shock by a factor of
      $\sim s^2$ (Fahr \& Chalov 2008; Zank et al. 2010; see equation
      \ref{eq:dpui} here), where $s \simeq 2.5$ is the compression
      ratio at the shock, and we take the preshock thermal velocity to
      be $ \ga 800 \km \s^{-1}$.  }}} For these parameters $G\simeq
0.1$ (Spitzer 1956).  These values bring the loss time scale to $t_s
\simeq 2000 \yr$, given the default values of the other parameters in
equation (\ref{eq:ts}). This is longer than the age of a PN at that
stage.

Adopting $t_s \sim 1000 \yr$, a PUI with $V_+ \simeq
2000 \km \s^{-1}$ travels a distance of $\sim 4 \times 10^5 \AU$.
Downstream (away from the shock), the gas in the bubble likely would be
further compressed, so this distance will be proportionally shorter.
Hence, within $\sim 100 \yr$ the PUIs can travel a distance of $\sim
10^5 \AU$.  If they are not deflected, they could reach the dense
nebular shell of a small PN (the visible shell where $T\sim 10^4 \K$),
where they will quickly decelerate ($t_s \simeq 1 \yr$) and deposit
their energy.  However, the deflections and tangled magnetic field
would likely prevent them from directly reaching the visible dense
shell on a straight trajectory.

As noted, in old PNs, where $r_s \ga 5000 \AU$, and ${\dot M_w} \la
{10^{-7} M_\odot \yr^{-1}}$, the PUI stopping time is larger than the
PN age, $t_s \ga 1000 \yr$, and the PUIs might efficiently carry
energy out of the hot bubble (if their number is large enough).
However, in younger PNs this is less likely.  For example, in
BD+30$^\circ$3639 the velocity is $\sim 700 \km \s^{-1}$ and the mass
loss rate is estimated to be ${\dot M_w} \sim 10^{-6} M_\odot \yr{-1}$
(Leuenhagen et al.\ 1996; Marcolino et al.\ 2007).  The outer hot
bubble radius is $\sim 5000 \AU$ and the shock radius $r_s$ is smaller
(Kastner 2008), so
the PUIs in the preshock region will lose their energy very quickly if
they are not formed very near the shock wave. On the other hand, in
this young PN, it is also possible that a large mass of dense clumps
would be formed by instabilities close to the shock wave, and that such clumps could supply
the require number density of PUIs to cool the newly-formed hot
bubble.

The main conclusion of this section is that if the PUIs originate
close to the center, at $r_o \la 10^3 \AU$, they will lose their
energy before entering the shock wave.  If PUIs are to play a role in
cooling the hot bubble, they must originate just inside the shock,
such that $r_o \la r_s$, with $r_o \ga 3\times 10^3 \AU$ (where the exact
value depends on the actual mass loss rate of the fast wind).  Hence,
PUIs might play a role in PNs that are not smaller than a few 10$^3
\AU$ as long as the fast-wind mass loss rate is also low, of the order
of ${\dot M_w} \la {10^{-7} M_\odot \yr^{-1}}$.  In such PNs, backflowing gas and
slow clumps can supply the required PUIs.

\section{SUMMARY}
\label{sec:summary}

Slowly moving ions that are picked up by the solar wind, called pick up
ions (PUIs), carry most of the energy in the post shock region
(Fig. \ref{fig:solar}); their presence explains the unexpectedly low
temperature of the shocked solar wind gas (e.g., Richardson et
a. 2008; Decker et al. 2008; Fahr \& Chalov 2008; Wu et al. 2009).
Motivated by these results for the solar wind, we examined whether a
similar process can occur in planetary nebulae (PNs). In PNs the flow
structure is similar, although not identical, to that of the solar
wind (Fig. \ref{fig:pn}). Hence, PUIs may also be present in PNs. If
so, the presence of PUIs might explain the general finding that the
temperature of the hot bubble formed by the post-shock gas in most PNs
(region FW2 in Fig. \ref{fig:pn}) is lower than that expected from
straightforward hydrodynamic shock calculations --- a discrepancy
(dubbed the {\it low temperature problem}) for which several
alternative explanations have been proposed but no clear consensus has
emerged (see section \ref{sec:intro}).

We demonstrate that the presence of PUIs might explain the PN low
temperature problem.  However, whereas in the case of the solar wind
the neutral atoms that turn into PUIs penetrate the pre-shock solar
wind region from the interstellar medium (ISM; see the schematic
particle trajectory drawn in Fig. \ref{fig:solar}), in PNs the
densities are much higher in all regions, and neutral atoms cannot
penetrate from regions outside the hot bubble so as to reach the
pre-shock region (region FW1 in Fig. \ref{fig:pn}).  Instead, we
hypothesize that, in PNs, the PUI source would most likely be slowly
moving clumps embedded in the fast wind or jets.  These clumps are
formed by instabilities or from backflowing cold gas, as discussed in
section \ref{subsec:source}.  For the PUIs behind the shock not to
thermalize too rapidly, the PUI stopping time (given by equation \ref{eq:ts})
cannot be shorter than the typical flow time. This
condition is met by a large margin for the solar wind, but only
marginally in PNs, and only under certain circumstances. In
particular, we find that the conditions under which PUIs might play a role in
moderating the hot bubble temperatures in PNs are (a) the slowly
moving clumps (the source of the PUIs) must be located just inside the
shock ($r_o \la r_s$, with $r_o \ga 3\times 10^3 \AU$), and (b) the mass
loss rate cannot be too large, i.e., ${\dot M_w} \la {10^{-7} M_\odot \yr^{-1}}$.

{{{  It is worth considering whether the fast wind itself
      would decelerate as a result of the incorporation of PUIs. To
      play any significant role in the fast wind shock, the PUI number
      fraction should be $\xi_{\rm P} \ga 0.1$ of the total fast wind
      particle number density.  As the typical thermal velocity of the
      preshock PUIs ($V_+$) is that of the bulk velocity of the wind,
      the bulk energy of the wind is reduced to $1-\xi_{\rm P}$ times
      its original value. This is a small change that cannot account
      by itself for the low temperature problem.  Even if we take this
      fraction to be $\xi_{\rm P}=0.3$, the reduction in energy is
      $\sim 30\%$, and the bulk wind speed is reduced by only $\sim 15
      \%$.  As other uncertainties in the winds interaction in PNs are
      larger, there is unlikely to be any significant reduction in the
      preshock velocity of the fast wind as a result of the PUI
      formation process.  }}}

Our results can apply for the case where the hot bubble are formed by
jets.  In these cases two opposite lobes are formed by two jets.
Namely, this process can occur in symbiotic nebulae.  Finally, it is
evident from our study that the comparison of the solar wind
termination shock with that of the fast winds in PNs has its own
scientific interest beyond the low temperature problem. Future studies
of this potential correspondence should shed further light on the
evolution of the shocked fast wind in PNs.

\acknowledgements {{{  We thank an anonymous referee for several clarifying comments
regarding PUIs in the solar wind. }}}
This research was supported by the Asher Fund for
Space Research at the Technion and a grant from the Israel Science
Foundation.  Support for J.K.'s research on planetary nebulae is
provided by NASA Astrophysics Data Analysis Program award NNX08AJ65G
to the Rochester Institute of Technology.

\end{document}